\documentclass[preprint,proceedings]{rmaa}

\suppressfulladdresses 

\usepackage{paralist}

\usepackage{psfrag,color}

\newcommand{\Msun}{~M_\odot}
\newcommand{\msun}{M_\odot}

\newcommand{\kms}{\rm ~km~s^{-1}}

\newcommand{\ml}{~\msun ~\rm yr^{-1}}

\SetYear{2006}
\SetConfTitle{Circumstellar Media and Late Stages of Massive Stellar
Evolution}

\title{Circumstellar Interaction Around Type Ib/c Supernovae
and the GRB Connection} 

\author{
  R. A. Chevalier\altaffilmark{1}}

\altaffiltext{1}{Department of Astronomy, University of Virginia,
P.O. Box 400325, Charlottesville, VA 22904, USA (rac5x@virginia.edu).}

\shortauthor{Chevalier}
\shorttitle{Type Ib/c Supernovae Circumstellar Interaction}

\listofauthors{R. A. Chevalier}
\indexauthor{Chevalier, R. A.}

\abstract{Radio observations of Type Ib/c supernovae suggest that
circumstellar interaction takes place with a wide range of wind
densities, comparable to that seen in Galactic Wolf-Rayet stars.
Efficient production of magnetic field in the shocked region is needed.
The X-ray emission observed from some Type Ib/c supernovae is higher
than would be expected by the thermal or inverse Compton mechanisms;
a synchrotron interpretation requires a flattening of the electron
energy spectrum at high energies, as might occur in a cosmic ray
dominated shock wave.
The wind density variations that are indicated in two supernovae may
be due to a binary companion, although variable mass loss from a single
star remains a possibility.
Other than the optical supernova radiation, the emission from the
nearby SN 2006aj/GRB 060218 appears to be powered by a central engine,
while that from SN 1998bw/GRB 980425 could be powered by either
a central engine or the outer supernova ejecta.
}

\addkeyword{Gamma rays: bursts}
\addkeyword{Stars: Mass loss}
\addkeyword{Supernovae: general}

\begin{document}
\maketitle

\section{Introduction}
\label{sec_intro}

Type I b/c supernovae (SNe Ib/c) have come to the fore of supernova research
because of their association with gamma-ray bursts (GRBs) and  
the fact that they
represent a significant mode of massive star death.
They are thought to be the explosion of massive stars that have lost all,
or nearly all, of their H envelopes; the SNe Ib have prominent He lines in
their spectra and the SNe Ic do not.
They explode as relatively compact Wolf-Rayet stars and thus have 
optical light curves that are dominated by power input from radioactivity.

Circumstellar interaction for these objects is of special interest because
of the extensive mass loss that has occurred leading up to the supernova
and because of GRB connection.
Here, I describe some of the general aspects
of the interaction (\S~\ref{sec_general}), discuss  events of special
interest (\S~\ref{sec_variations}), consider the relation between SNe Ib/c and
apparently
low energy GRBs (\S~\ref{sec_grb}), and summarize the issues (\S~\ref{sec_disc}).

\section{General Aspects of Circumstellar Emission}
\label{sec_general}

The most extensive observations of circumstellar interaction are
at radio wavelengths (Sramek et al. 2005).
The general picture of the interaction is that in the initial stellar
explosion, the shock wave accelerates through the outer part of the
star leading to a steep power law density profile in the freely
expanding ejecta.
The maximum velocity that is achieved is partly determined by the
radius of the progenitor star because the shock acceleration stops
when the radiation is able to freely stream away at the time of shock
breakout.
A more compact star has greater shock acceleration because there is a
larger ratio between the average stellar density and the density at the
photosphere.
Once the supernova shock has broken out of the star and
a shock wave develops in the surrounding wind, a shocked layer forms
bounded by a reverse shock on the inside and a forward shock on the outside.
The shock waves give rise to particle acceleration and radio synchrotron
emission.

At early times, there is absorption of the radio emission by synchrotron
self-absorption (SSA); additional processes, such as free-free absorption by
unshocked wind gas ahead of the forward shock wave, may also play a role.
In the case of SNe Ib/c, the circumstellar density is relatively low because
of the high progenitor wind velocity and the shock velocity is high,
which favors SSA as the dominant process (Chevalier 1998).
Observations of radio light curves support the hypothesis that it is
dominant in those cases where detailed observations are available
(Soderberg et al. 2005, 2006).
Fig. 1 shows the positions of well observed radio supernovae in a plot of
peak radio luminosity vs. time of the peak of the radio light curve.
If the dominant absorption process is SSA, application of synchrotron
theory allows the radius of the radio emitting region, and thus the velocity,
to be found.
There is the assumption of energy equipartition between the magnetic field and
the relativistic electrons, but the results are insensitive to this assumption.

\begin{figure}[!t]
  \includegraphics[width=\columnwidth]{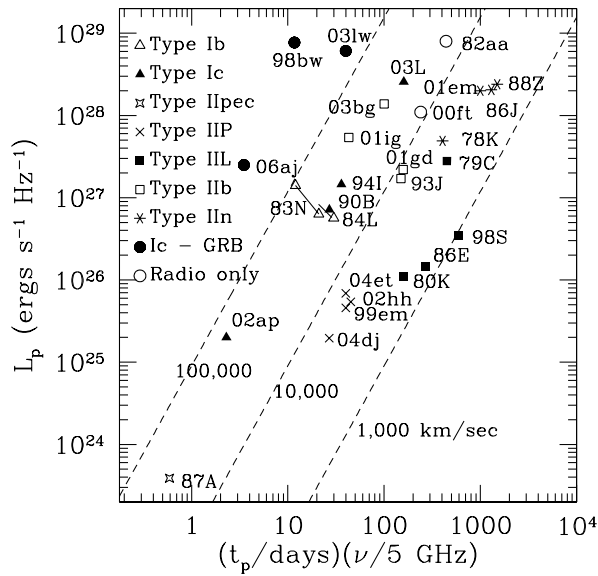}
  \caption{Peak luminosity and corresponding epoch for the
well-observed radio supernovae. The dashed lines give curves of constant
expansion velocity, {\it assuming} synchrotron self-absorption 
at early times (updated version of Fig. 4 in Chevalier 1998).}
  \label{fig:simple}
\end{figure}

Fig. 1 shows that the typical velocity of the SNe Ib/c near maximum radio
luminosity is $\sim30,000\kms$, which is systematically higher that the
velocities present in SNe II.
There are several reasons for this difference.
First, there are higher initial velocities in the SNe Ib/c because of their
more compact progenitors.
The maximum velocity can be $\sim100,000\kms$ for typical
supernova parameters.
In the case of SNe IIP (plateau light curve), with red supergiant progenitors,
the maximum velocity may be $\sim15,000\kms$.
Second, the Wolf-Rayet star progenitors of SNe Ib/c have winds with a high
typical velocity, $\sim1000\kms$, leading to a low circumstellar density.
There is thus relatively less deceleration during the circumstellar interaction.
Finally, the strong mass loss during the stellar evolution can lead to a relatively
low mass at the time of the SN Ib/c explosion.
The high energy-to-mass ratio leads to a higher mean velocity for the ejecta
and thus to a high interaction velocity.

Fig. 1 also shows that there is a large range in radio luminosities of the
the SNe Ib/c.
In the case of a radio light curve with SSA, there is not sufficient information to
deduce the physical parameters for the interaction, but Chevalier \& Fransson
(2006) noted that the range in luminosity roughly corresponds to the range of 
circumstellar density inferred for Galactic Wolf-Rayet stars if the efficiency
of production of magnetic fields and relativistic electrons is fairly high
($\epsilon_\mathrm{e}\approx\epsilon_\mathrm{B}\approx 0.1$).
Here, the $\epsilon$'s are the ratios of energy densities in particles and fields
to $\rho_0 v_\mathrm{s}^2$, where $\rho_0$ and $v_\mathrm{s}$ are the preshock density and velocity
of the forward shock wave.
The magnetic field must be amplified in the interaction region and cannot just
be the compressed wind field (Chevalier 1998).
One possibility for amplification is the hydrodynamic instabilities that
occur near the contact discontinuity (Jun \& Norman 1996);
however, it is not clear that this can achieve the required efficiency.
Another possibility is that there are instabilities associated with efficient
cosmic ray acceleration that can increase the magnetic field (Bell 2004).
This mechanism has the property that it is more efficient at higher shock
velocities, which would explain why the SNe Ib/c have more efficient
production of synchrotron radiation than do Galactic supernova remnants.

X-ray emission has been detected from a number of SNe Ib/c, but there
are not extensive light curves as there are for some 
at radio wavelengths.
In general, the required emission cannot be thermal unless the circumstellar
density is considerably higher than expected around a Wolf-Rayet star.
Although this scenario cannot be completely ruled out, it is more plausible
that the radiation mechanism is nonthermal.
The possibilities are inverse Compton emission and synchrotron radiation.
Inverse Compton emission involves the scattering of photospheric photons with
the relativistic electrons in the interaction region and is thus most likely
to be important near optical maximum light.
The Type Ic SN 2002ap has an early observation from {\it XMM} that can be 
interpreted in terms of inverse Compton emission (Bj\"ornsson \& Fransson 2004;
Sutaria et al. 2003).
However, in most cases the X-ray observations are at a time when the optical
light is not strong and the inverse Compton mechanism is not viable.
This is especially true for {\it Chandra} observations of SN 1994I at an
age of $6-7$ yr (Immler et al. 2002).
In most cases, the extension of the observed radio synchrotron spectrum to
X-ray energies falls below the observed X-ray fluxes.
Chevalier \& Fransson (2006) suggested that the relativistic electron
spectrum produced in a cosmic ray dominated shock could have a flattening at
high energies (e.g., Ellison et al. 2000) that allows 
the observed X-ray fluxes to be reproduced
in a synchrotron model.
More detailed observations are needed to check on this hypothesis.

For the velocities and densities that are present in the circumstellar 
interaction of SNe Ib/c, radiative cooling is not important for the shocked
gas so that there is no cool gas present that could give optical emission.
There is the possibility that the X-ray emission irradiates the ejecta,
giving rise to observable effects, but that has yet to be established.

\section{Mass Loss Variations}
\label{sec_variations}

While the general properties of the radio supernovae are consistent
with interaction with a surrounding wind, there are some cases
where there is evidence for structure within the wind.

\subsection{SN 2001ig and SN 2003bg}

As can be seen in Fig. 1, SN 2001ig
(Ryder et al. 2004) and SN 2003bg 
(Soderberg et al. 2006) have radio properties and
inferred velocities that are in line with those expected for SNe Ib/c.
As noted above, the high velocities are typical for those the explosions
of relatively compact, Wolf-Rayet stars.
The optical light curve of SN 2003bg (Hamuy, priv. comm.) 
was consistent with the explosion of
a relatively compact star.
However, both SN 2001ig and SN 2003bg showed H lines in their spectra
(Ryder et al. 2004; Hamuy et al. 2003), although SN 2003bg
was initially classified as a Type Ic supernova 
of the broad-lined type (Filippenko \& Chornock 2003).
The H lines observed in SN 2003bg were very broad.
Apparently, the supernovae exploded as Wolf-Rayet stars with a small amount of H,
an amount insufficient to produce an extended envelope
($\la 0.02\Msun$).

Ryder et al. (2004) classified SN 2001ig as a Type IIb supernova, which
is the same as SN 1993J.
However, SN 1993J exploded as a red supergiant, with $\sim0.1-0.2\Msun$ of
H at the time of the explosion.
Compared to the other 2 supernovae, SN 1993J showed lower expansion velocities
of the radio emitting region (Fig. 1) and the interaction with the dense
red supergiant wind produced a radiative reverse shock wave with strong
optical signatures.
A preferred explanation for the mass loss from the SN 1993J progenitor was
a close binary companion (e.g., Woosley et al. 1994), 
and a likely binary companion was eventually
found (Maund et al. 2004).
The binary separation is determined by the fact that the binary
interaction must drive mass loss when the progenitor star is a red
supergiant; in the models of Woosley et al. (1994), the final
separation is $6.4-14.3$ a.u. or $(0.96-2.1)\times 10^{14}$ cm.
The corresponding range of binary periods is $5-15$ yr.

SN 2001ig and SN 2003bg were similar 
to each other not only in their basic radio properties,
but also in detailed properties of their radio light curves: both showed a
flux increase at around the time they were turning optically thin and later
showed a flat section of light curve (Ryder et al. 2004; Soderberg et al. 2006).
In the context of the interaction of 
freely expanding supernova ejecta with a surrounding wind,
it is not plausible that the light curve features are due to structure in
the supernova ejecta.
The largest effect would be approximately 
constant velocity of the interaction region produced by a steep density
gradient, but this would produce little change in the radio light curve
compared to the normal case of slow deceleration.

The most likely reason for the flux variations is thus density variations
in the circumstellar medium.
A previous model of this type was for the regular flux variations in the radio light
curve of SN 1979C over a time of 10 yr (Weiler et al. 1992);
the amplitude of the variations was $\sim20$\%.
Weiler et al. (1992) attributed the variations to stellar
pulsations or, more likely, to a binary companion.
The binary would create a spiral pattern in the slow wind; Schwarz \&
Pringle (1996) showed how a binary companion with a velocity comparable to
the wind velocity would create structure within the slow wind by gravitational
effects.
If there is a continual spiral structure and the emitting region is optically thin,
the effect on the radio light curve should be steady and regular features in
the light curve would not be expected.
Thus, the suggestion is that the binary orbit is eccentric so that there is
asymmetric structure in the spiral pattern (Weiler et al. 1992; 
Schwarz \& Pringle 1996).
The required binary period is $t_\mathrm{b}\approx t_\mathrm{per}
v_\mathrm{sn}/v_\mathrm{w}$, where $t_\mathrm{per}$ is
the timescale for the periodic structure in the light curve, $v_\mathrm{sn}$ is
the average velocity of the supernova shock, and $v_\mathrm{w}$ is the wind velocity
of the progenitor star.
In the case of SN 1979C, $t_\mathrm{per}=1575$ d (Weiler et al. 1992), 
$v_\mathrm{sn}\approx 10,000\kms$,
and $v_\mathrm{w}\approx 10\kms$, so that $t_\mathrm{b}\approx 4300$ yr.

In the case of SN 2001ig, Ryder et al. (2004) also appealed to a companion
in an eccentric orbit to produce the density structure.
Ryder et al. (2006) in fact found a supergiant 
star at the position of the supernova
that can plausibly be identified as the progenitor companion.
The fact that SN 2001ig and SN 2003bg require similar binary 
parameters initially seems
surprising, but becomes more plausible considering that the binary may play
a crucial role in producing the mass loss leading to the Type IIb supernovae,
as in the case of SN 1993J.
The binary period needed for the flux variations is $t_\mathrm{b}\approx 13$ yr, based
on $t_\mathrm{per}=160$ d, $v_\mathrm{sn}\approx 30,000\kms$,
and $v_\mathrm{w}\approx 1000\kms$.
The approximate agreement between this binary period and
the binary period needed to produce a SN 1993J type system is a point in
favor of the binary hypothesis.
If the binary in SN 1993J caused structure in the wind, the effect
on the radio light curve would have a rapid timescale (days) because
of the slow red supergiant wind velocity.
On this timescale, asymmetric structure and inhomogeneities  would
be likely to wash out radio flux variations; the radio light
curves of SN 1993J are, in fact, smooth
(Van Dyk et al. 2005).

In the models for SN 1993J
(Woosley et al. 1994), the binary interaction leaves
$\sim0.1-0.2\Msun$ of envelope material on the supernova progenitor
so there is the question of how SN 2001ig and SN 2003bg ended up
with considerably less.
At the time of explosion, the SN 1993J progenitor is thought to have been
losing $\sim(3-4)\times 10^{-5}\ml$ (for a wind velocity of $10\kms$)
(Fransson et al. 1996), so if
up to $10^4$ yr pass between the binary interaction and the supernova,
there is the possibility of most of the remaining H envelope being lost.

A possible problem is that if the binary goes through a phase where the binary
separation is comparable to the stellar radius (of the red supergiant),
the binary orbit should be circularized by tidal effects, 
giving rise to the question of
how the asymmetric structure is created for the radio variations.
A possibility is that, at early times, the interaction with only the spiral
pattern in front of the supernova can be observed, because the radio emission
from the back side is free-free absorbed by the supernova ejecta.
In this case, the flux variations should become less marked with time as
optical depth effects become less important.
In fact, for both SN 2001ig and SN 2003bg the initial flux increase at
$\sim160$ d is approximately a factor of 2, whereas the next feature is
simply a flattening of the light curve.
These properties can be contrasted with the roughly steady undulations in
the radio light curve of SN 1979C (Weiler et al. 1992).

Another issue is the physical mechanism for the regular density structure.
In the scenario of gravitational effects discussed by Schwarz \& Pringle (1996),
the gravitational radius ($2GM_\mathrm{com}/v_\mathrm{w}^2$ where $M_\mathrm{com}$ is the
companion star mass) is large because $v_\mathrm{w}=10\kms$.
When $v_\mathrm{w}=1000\kms$, as is the case here, the gravitational radius is
much smaller and only a small fraction of the progenitor wind is affected
by the companion star.
Ryder et al. (2004) note that spiral dust patterns have been observed
around some Wolf-Rayet stars as a result of binary interaction
(e.g., Tuthill et al. 1999, 2006) and suggest that such a structure
results in the radio light curve variations.
The initial cases of this phenomenon that were discovered had a
binary period $<1$ yr, but Tuthill et al. (2006) find evidence for
periods up to $\sim2.5$ yr.
However, this spiral structure is the likely result of wind-wind interaction
and the companion star must be an O 
or early B star to have a strong enough wind
to produce structure in the Wolf-Rayet wind.
In the case of SN 2001ig, the apparent companion star is a 
late B through early F supergiant 
(Ryder et al. 2006) and it is questionable whether it has a sufficiently
strong wind to produce the needed structure.

There is thus some degree of uncertainty in the binary hypothesis and
single star possibilities should also be considered.
Kotak \& Vink (2006) suggested that the progenitor is an S Doradus
star, which is related to the  luminous
blue variables (LBVs).
The lower wind velocity in this case, $100-500\kms$, would give a
longer timescale for the structure in the wind, $26-100$ yr.
Kotak \& Vink (2006) suggest that this timescale is compatible
with that observed during a ``long-S Doradus'' phase.

The two possibilities of S Dor or Wolf-Rayet star present
progenitors with different properties.
The properties of S Dor stars discussed by van Genderen (2001)
indicate a radius range of $3\times 10^{12}-2\times 10^{13}$ cm,
much larger than a Wolf-Rayet star.
Also, to have the requisite mass loss from a single star would
require an initial mass $\ga30\Msun$, whereas a lower mass is
possible if there is mass loss driven by binary interaction.
As mentioned above, the high 
expansion velocity indicated by the radio emission
(Fig. 1) implies a fairly compact progenitor, although some part
of the S Dor range may be permitted.
A better way to distinguish between the possibilities would
be to compare information on the light curves and spectra of
the supernovae with models for the explosions.
This has yet to be carried out.

\subsection{SN 2001em}

SN 2001em was initially discovered as a Type Ib or Ic supernova,
with Ic considered more likely (Filippenko \& Chornock 2001).
It was detected as a luminous radio source at an age of 767 days
(Stockdale et al. 2004), which led to a detection of X-ray emission
(Pooley \& Lewin 2004) and the finding of a relatively narrow 
H$\alpha$ line (Soderberg et al. 2004).
The radio luminosity for the age is shown in Fig. 1, which shows
that its luminosity is comparable to those of the most luminous
Type IIn supernovae at a comparable age.
This supernova designation is also indicated by H$\alpha$ line.
The implication is that the supernova was transformed from a Type Ib/c,
with relatively little circumstellar interaction, to a Type IIn, with
strong interaction.

The large radiated luminosity implies that a significant fraction of
the supernova energy (assuming it to be $\sim10^{51}$ ergs) was
thermalized, so that the supernova ejecta must interact with a surrounding
mass that is comparable to the ejecta mass.
The timescale for the interaction, coupled with the mean supernova
velocity, suggests a radial scale $\sim10^{17}$ cm (Chugai \& Chevalier 2006).
The presupernova mass loss timescale is then $\sim3000(v_\mathrm{w}/10\kms)^{-1}$
 yr and the
corresponding mass loss rate is $\sim10^{-3}(v_\mathrm{w}/10{\kms})\ml$.
Such a high rate of mass loss is unusual.
One possible Galactic example is the eruptive phases of LBVs
(Smith 2006).

\section{The Connection to Low Luminosity GRBs}
\label{sec_grb}

Three Type Ic supernovae have been found in association with apparently
low luminosity GRBs: SN 1998bw with GRB 980425 (Galama et al. 1998), 
SN 2003lw with GRB 031203 (Malesani et al. 2004),
and SN 2006aj with GRB 060218 (Campana et al. 2006).
The relative proximity of these events provides an excellent opportunity
to examine the relationship between supernovae and GRBs.
In all of these cases, the optical emission after about
a day was dominated by the light
from the supernova.
SN 2006aj was also observed at very early times
(Campana et al. 2006), when the issue is not clear;
Campana et al. (2006) interpreted the early optical emission as shock
breakout emission from the supernova.
All 3 events were observed as radio and X-ray sources.
It can be seen in Fig. 1 that the 3 objects have radio properties that
imply rapid (approximately relativistic) expansion of the radio
emitting region.
An important question is whether the emission can be attributed to
supernova expansion and shock waves, as in the case of normal SNe Ib/c,
or to the activity of a central engine, as is needed for a GRB.

\subsection{SN 1998bw and GRB 980425}

An especially good data set is the radio data set on SN 1998bw
(Kulkarni et al. 1998).
Li \& Chevalier (1999) modeled the radio light curves and claimed
that the emission was most likely related to a central engine and
not to the outer supernova ejecta.
There were several reasons for the claim:
(i) there is insufficient energy at high velocity in the supernova
to produce the radio emission; (ii) the flux rise observed around
days $20-35$ requires the action of the central engine; and (iii) outside
of the time of flux increase the evolution is consistent with a constant
energy explosion, which is typical of GRB afterglows.

On (i), it is clear from \S~\ref{sec_general} that a supernova
would not be able to produce the emission if the energy were
comparable to that of a normal supernova ($\sim10^{51}$ ergs).
In this case, the shock acceleration would not attain sufficiently
high velocities at the outer edge of the star to explain the
radio emission (Li \& Chevalier 1999).
However, there have been estimates that the energy in SN 1998bw
was as high as $5\times 10^{52}$ ergs in models that fit
the early evolution (Nakamura et al.  2001).
With the high energy, the breakout velocities can become
relativistic and the density at a particular velocity in the outer
steep part of the density profile is $\propto E^{3.59}$
(Chevalier \& Fransson 2006).
Tan et al. (2001) have carried out detailed calculations of the
transrelativistic expansion expected for a high energy explosion,
finding that the energetics required by the radio emission can
be satisfied.
Thus, the reason for the appearance of supernovae in the high
velocity part of Fig. 1 could be an usually high supernova energy.

The argument for the flux increase being due to a central engine
as opposed to a circumstellar density feature was the behavior of
the flux in the optically thick regime.
However, that argument cannot be considered to be conclusive;
the flux rise has a similar appearance to those seen in SN 2001ig
and SN 2003bg.
In the case of SN 1998bw, the shock radius when the flux rise
occurs is $\sim(1-1.5)\times 10^{17}$ cm
(Li \& Chevalier 1999).  
The inferred radius is $2-3$ times larger than in the case of the
other supernovae; this could simply reflect the time when the
effect of the density structure was most noticeable.
A difference in the objects is that SN 2001ig and SN 2003bg showed
H lines in their spectra, whereas models for SN 1998bw indicate
the explosion of the bare C/O core of a massive star (e.g.,
Nakamura et al. 2001).
However, there have been suggestions that SN 1998bw-like explosions
could contain some H (Branch et al. 2006).

The final argument involves the rate of decline of the radio emission,
outside of the flux increase episode.
Li \& Chevalier (1999) took the relatively rapid decline to be
indicative of a constant energy explosion, which is the typical
assumption for GRB afterglows.
However, assumptions about the evolution of the efficiency of
magnetic field and relativistic electron production play a role in
this interpretation, so the case is uncertain.
Overall, there is not compelling evidence for either the central
engine or supernova origin of the radio emission.

Over days $10-200$ (including the time of the flux
increase), the optically thin radio flux from SN 1998bw
declined by a factor of $\sim50$, while the X-ray evolution was
essentially flat (Pian et al. 2000; Kouveliotou et al. 2004).
Waxman (2004) modeled the early X-ray emission as synchrotron emission
from a shocked region driven by a supernova mass shell moving
at constant velocity ($0.8c$); the shell mass was $\sim1\times 10^{29}$ gm.
Waxman did not discuss the origin of the shell, but a shell can
be formed at the time of supernova shock breakout because of the
escape of the radiative energy.
The shell mass is $M_\mathrm{s}=4\pi R^2\Delta R \rho=4\pi R^2\tau /\kappa$,
where $R$ is the radius at shock breakout, $\Delta R$ is the thickness
of the region that collapses,  $\rho$ is the density, $\tau\approx c/3v$
is the optical depth at which radiation begins to leak out, $v$ is
the shock velocity, and $\kappa$ is the opacity (Chevalier 1981).
Allowing for shock breakout in an optically thick wind at
$R=2\times 10^{12}$ cm and opacity due to electron scattering, the
shell mass is still orders of magnitude smaller than that needed
for SN 1998bw.
The implication is that the wind interaction is with the outer
steep power law part of the supernova density distribution.
The interaction shell produced in this way decelerates slowly, so
there is not a large change from the constant velocity case;
however, the steepening of the X-ray light curve at $100-200$ day,
attributed by Waxman (2004) to the deceleration of the mass shell,
is not produced in a natural way.
However, the evidence for a break depends on just one
X-ray point, that at day 200.

In the model of Waxman (2004), the radio and X-ray emission are both
synchrotron emission from the interaction shell.
Because the cooling frequency, $\nu_\mathrm{c}$, is between radio and X-ray
wavelengths, the radio flux evolves as $t^{-1/2}$ while the X-ray
evolution is flat.
The observed radio decline is steeper than this.
In addition, Waxman (2004) assumes a relativistic electron energy
spectral index $p=2$, whereas the radio observations imply that
$p=2.5$ (Kulkarni et al. 1998; Li \& Chevalier 1999).
An extrapolation of the radio spectrum on day 100 falls below
the observed X-ray flux, even when the steepening of the spectrum
due to synchrotron cooling is not taken into account.
If the radio and X-ray emission are from the same region, there
must be flattening of the particle spectrum toward high energies.
Chevalier \& Fransson (2006) discussed how this could be important
for ordinary SNe Ib/c; the mechanism for the flattening was
particle acceleration in a cosmic ray dominated shock.
The mechanism needs to be examined for the case of a higher shock
velocity, as is present in SN 1998bw.

\subsection{SN 2006aj and GRB 060218}

Fig. 1 shows that the 3 SNe Ic associated with GRBs have radio
properties implying unusually high velocities.
In the case of SN 1998bw, the general radio properties could be
attributed to a high energy explosion or to a central engine.
However, SN 2006aj was observed to have optical properties
similar to SN 2002ap, which was not very energetic.  From 
modeling the light curve and spectra of SN 2006aj,
Mazzali et al. (2006) estimate an energy of $E\sim2\times 10^{51}$ ergs
and an ejecta mass of $M_\mathrm{e}\sim2\Msun$; these values can be compared
to $4\times 10^{51}$ ergs and $3\Msun$ deduced for SN 2002ap
(Tomita et al. 2006), which seems to have had little
high velocity ejecta (Berger et al. 2002).
On the basis of the radio emission from SN 2006aj,
Soderberg et al. (2006) deduced a minimum energy of $(1-2)\times
10^{48}$ ergs in ejecta with a Lorentz factor of 2.3 (velocity
of $270,000\kms$).
For the above values of $E$ and $M_\mathrm{e}$ and the density profile for
a SN Ib/c given in Berger et al. (2002), the energy above a
velocity of $200,000\kms$ is $\sim10^{46}$ ergs.
The implication is that the radio emission is related to
the central engine that also produced the GRB.

Taking $F_{\nu}\propto \nu^{\beta}t^{\alpha}$ gives $\alpha=-0.8$
and $\beta=-0.55$ for the radio emission from
SN 2006aj (Soderberg et al. 2006).
Except for SN 2002ap, the normal SNe Ib/c have $\alpha=-(1.2-2)$, with
an average of $-1.5$; the typical value of $\beta$ is $-1$
(Chevalier \& Fransson 2006).
SN 2002ap, which is the normal SN Ib/c with the highest velocities has
$\alpha=-0.8$, although this may be due to the importance of
inverse Compton losses for the radiating particles (Bj\"ornsson \& Fransson
2004).
In the case of SN 2006aj, estimates of inverse Compton effects indicate
that they are not important, which is supported by  the relatively
flat spectral index.
In fitting a standard GRB afterglow model to the radio emission from
SN 2006aj, Soderberg et al. (2006) found that the best model was
expansion into a constant density medium with $n=5$ cm$^{-3}$.
This result, which is typical of GRB afterglow models, is at odds
with the expected wind medium expected around a Wolf-Rayet star and
with  the medium deduced for normal SNe Ib/c.

The X-ray emission from SN 2006aj, with a steep spectrum $\beta=-2$,
cannot be considered an extension of the radio synchrotron emission in
any straightforward way (Soderberg et al. 2006).
The steep spectrum, together with the moderate rate of decline ($\alpha=-1$),
is suggestive of continued power input, and Soderberg et al. (2006)
suggest that the power is provided by a central magnetar.

An additional component in SN 2006aj/GRB 060218 is an early thermal X-ray
component, which Campana et al. (2006) interpreted as the shock breakout
emission from the supernova.
However, the radiated energy in this component, $\ga 10^{49}$ ergs
(Campana et al. 2006; Li 2006), is greater than that produced
in a plausible Wolf-Rayet explosion model by a factor $\ga 100$
(Li 2006).
In addition, the duration of the thermal component, $\sim2700$ s, is
larger than can be accomodated in a Wolf-Rayet explosion model,
even allowing for an emitting region out in an optically thick
progenitor wind (Campana et al. 2006; Li 2006).
The implication is that the thermal component is formed by a
flow from a central engine, as well as the nonthermal emission.

\section{Discussion}
\label{sec_disc}

The radio emission from normal SNe Ib/c is consistent with
synchrotron radiation produced by the outer supernova ejecta
interacting with the wind from the progenitor star.
Early absorption is generally due to synchrotron self-absorption.
Consistency with the wind densities expected around Wolf-Rayet
stars requires a high efficiency of production of relativistic
electrons and magnetic fields.
Hydrodynamic instabilities in the decelerating interaction region
may not be sufficient to produce the magnetic field and
shock wave instabilities might play a role.
The X-ray emission from SNe Ib/c requires a nonthermal mechanism
if the surrounding densities are typical of the winds from
Wolf-Rayet stars.
A synchrotron origin for late X-ray emission requires a flattening
of the particle spectrum to high energies, as might occur as a result
of particle acceleration in a cosmic ray dominated shock wave.

Beyond the standard wind interaction model, there are deviations
from the expected smooth radio evolution in some SNe Ib/c.
In SN 2001ig and SN 2003bg, there are density variations that
might be produced by a binary companion that is also responsible
for loss of the H envelope of the progenitor star.
A more dramatic radio (and X-ray) brightening occurred in SN 2001em,
indicating that the exploding star was interacting with a substantial
part of the lost H envelope.
Such extreme mass loss is rare for Galactic stars; one example is
perhaps the eruptive phase of luminous blue variables.

There is good evidence that long duration GRBs are associated with
SNe Ib/c, but it has not been possible to relate the interaction
properties to the circumstellar environment of Wolf-Rayet stars,
as in the case of normal SNe Ib/c.
Even for the nearby, low luminosity GRBs associated with well-studied
supernovae, the type of circumstellar interaction is unclear.
In the case of SN 1998bw/GRB 980425, there is uncertainty whether
the radio and X-ray emission is powered by the outer supernova
ejecta or by an inner engine as in GRBs.
In the case of SN 2006aj/GRB 060218, except for the optical
supernova emission the properties are most likely due to 
a relativistic flow from a central engine.

\acknowledgements
I am grateful to Nikolai Chugai, Claes Fransson, Zhi-Yun Li, and Alicia
Soderberg for their collaboration on various aspects of
this research.
The research was supported in part by NASA grant NNG06GJ33G
and NSF grant AST-0307366.

\end{document}